%
%
%

\ifx\mnmacrosloaded\undefined \input mn\fi
\input psfig

\overfullrule=0.0pt
\def\Tg{T_\gamma}
\def\barra{\biggr\vert}
\def\ref{\bibitem}


\pageoffset{-2.5pc}{0pc}

%
  
%
%
%



\pagerange{0--0}    
\pubyear{0000}
\volume{000}

\begintopmatter  

\title{Time--dependent analysis of spherical accretion onto black holes}

\author{L.~Zampieri,$^1$ J.~C.~Miller,$^{1,2}$ and R.~Turolla$^3$}
\affiliation{$^1$ International School for Advanced Studies (S.I.S.S.A.),
Via Beirut 2--4, 34013 Trieste, Italy}
\affiliation{$^2$ Nuclear and Astrophysics Laboratory, University of Oxford,
Keble Road, Oxford OX1 3RH, England}
\affiliation{$^3$ Department of Physics, University of Padova, Via Marzolo 8,
35131 Padova, Italy}

\shortauthor{L. Zampieri, J. C. Miller and R. Turolla}

\shorttitle{Time--dependent analysis of spherical accretion}


\acceptedline{Accepted ... Received ... ;
  in original form 1995 June 5}

\abstract {
Results are presented from a time--dependent, numerical investigation of
spherical accretion onto black holes, within the framework of relativistic
radiation hydrodynamics. We have studied the stability of self--consistent,
stationary solutions of black hole accretion with respect to thermal and
radiative perturbations and also the non--linear evolution of unstable, high
temperature models, heated by the hard radiation produced by the
accretion flow itself in the inner region near to the horizon. In some cases,
a hydrodynamic shock forms at around $10^3$--$10^4$
Schwarzschild radii, where Compton heating exceeds radiative cooling. The
calculations were made using a suitably designed radiation hydrodynamics code,
in which radiative transfer is handled by means of the PSTF moment formalism
and which contains an original treatment of the radiation temperature
equation.}


\maketitle  

\section{Introduction}

Stationary, spherical accretion onto black holes is a well--known and
extensively studied topic. Starting from the seminal paper by Bondi (1952),
many papers have been devoted to the analysis of spherical accretion under a
variety of conditions, mainly in order to obtain a definite estimate of the
efficiency of the process. In contrast with accretion onto neutron
stars, the efficiency is not fixed by the requirement that all the
kinetic energy of the accreting gas must be converted into radiation,
since no rigid surface exists which can stop the flow. Matter can cross
the horizon carrying a substantial fraction of the gravitational potential
energy liberated and the efficiency of the process is determined solely
by the effectiveness of the radiative processes in converting the internal
energy of the accreting gas into radiation, as first noted by Shvartsman
(1971). As shown by many authors (see e.g. Michel 1972; Novikov \&
Thorne 1973; Blumenthal \& Mathews 1976;
Begelman 1978; Brinkmann 1980), the flow properties are fixed
once the accretion rate is specified, so that stationary solutions
can be completely characterized by their position in the (${\dot m}$, $l$)
plane, where ${\dot m}$ and $l$ are, respectively, the accretion rate
and luminosity in Eddington units. Stationary spherical accretion
onto black holes was investigated in detail by Nobili, Turolla \& Zampieri
(1991, hereafter NTZ); Fig.~1 shows the (${\dot m}$, $l$) diagram for the
complete set of their solutions. At low accretion rates (${\dot m} < 1$),
spherical accretion is very inefficient in converting gravitational energy
into radiation since the density is too low for cooling processes to be
effective; the emitted luminosity is also very low (lower branch in Fig.~1,
hereafter the LL branch).
These models are essentially adiabatic and have very high temperatures
(see also Shapiro 1973a). In this regime, the only possibility for increasing
the efficiency of the accretion process is related to the presence of magnetic
fields, which can cause strong dissipation (e.g. through reconnection of field
lines) and induce strong emission of synchrotron radiation
(Shapiro 1973b; M\`esz\`aros 1975). Soffel (1982) studied in some detail
the transition from the optically thin regime to the optically thick one:
as ${\dot m}$ increases, the cooling processes become more effective
and the gas temperature decreases, causing in turn a decrease
in the total emitted luminosity (with a local minimum at around ${\dot m}
\simeq 0.1$). For higher values of the accretion rate, free--free absorption
is no longer negligible and the gas becomes optically thick in the inner
region near to the horizon of the black hole. The temperature increases because
heating exceeds cooling and also the luminosity rises since radiation is
in LTE with the gas in the inner core. Preliminary investigations of
spherical accretion in the diffusion regime were made by Tamazawa
et al. (1974), Maraschi, Reina \& Treves (1974),
Kafka \& M\'esz\'aros (1976), Vitello (1978),
Begelman (1979), Gillman \& Stellingwerf (1980) and Freihoffer (1981),
while a complete treatment was finally given by Flammang (1982, 1984),
who showed the existence of a subcritical point related to the equation
for the radiative luminosity.
When ${\dot m} > 1$, the inner core
starts to be optically thick to electron scattering as well. In these
conditions, a trapping radius appears (Rees 1978; Begelman 1978), below
which photons are
advected into the black hole, since their outward diffusion velocity
is smaller than the inward velocity of the accretion flow. This makes
the process less efficient and the rate of increase of luminosity with ${\dot
m}$ becomes slower. The stationary, hypercritical regime has been investigated
by Blondin (1986).

For $3 \la {\dot m} \la 100$, there is also another class of solutions,
characterized by having high temperatures and luminosities
(upper branch in Fig.~1, hereafter the HL branch).
These are dominated by the effects of comptonization which keeps the gas
and radiation temperatures almost equal in the inner part of the flow where
the density is sufficiently high to make the Compton parameter very large.
In the intermediate region between $10^2$ and $10^5 \, r_g$
(where $r_g$ is the Schwarzschild radius of the black hole), Compton heating
dominates and the only competitive cooling mechanism is free--free emission.
The first authors to investigate the possible existence of high
luminosity solutions were Wandel, Yahil \& Milgrom (1984) and Park
(1990a,b), who performed a detailed study of spherical accretion
for a large range of accretion rates and considered also a
two--temperature model with pair production.
High luminosity stationary solutions have relatively high efficiency
and appear to exist only for a very definite range of accretion rates.
Already in 1976, Ostriker and collaborators (Ostriker et al. 1976) pointed out
that, because of the non--local nature of comptonization, the heating produced
in the flow by the hard radiation coming from the inner region
can increase the gas temperature in such a way that the internal energy density
becomes larger than the gravitational energy density and the accretion process
is then stopped. This effect is called {\it preheating}. Later on, Cowie et al.
(1978), Shull (1979), Stellingwerf \& Buff (1982) and Krolik \& London (1983)
showed that preheating is very important in placing limits on the region of
parameter space within which HL solutions for black hole accretion can exist,
although the strength of preheating
is reduced if Compton cooling is taken into account (see e.g.
Bisnovatyi--Kogan \& Blinnikov 1980).
As shown by NTZ, preheating {\it at} the sonic point for the matter prevents
the existence of high luminosity solutions with ${\dot m} \la 3$, while
preheating {\it within} the sonic radius prevents the existence of
stationary solutions for ${\dot m} \ga 100$.
\beginfigure*{1}
\hskip 1.5 truecm
\psfig{figure=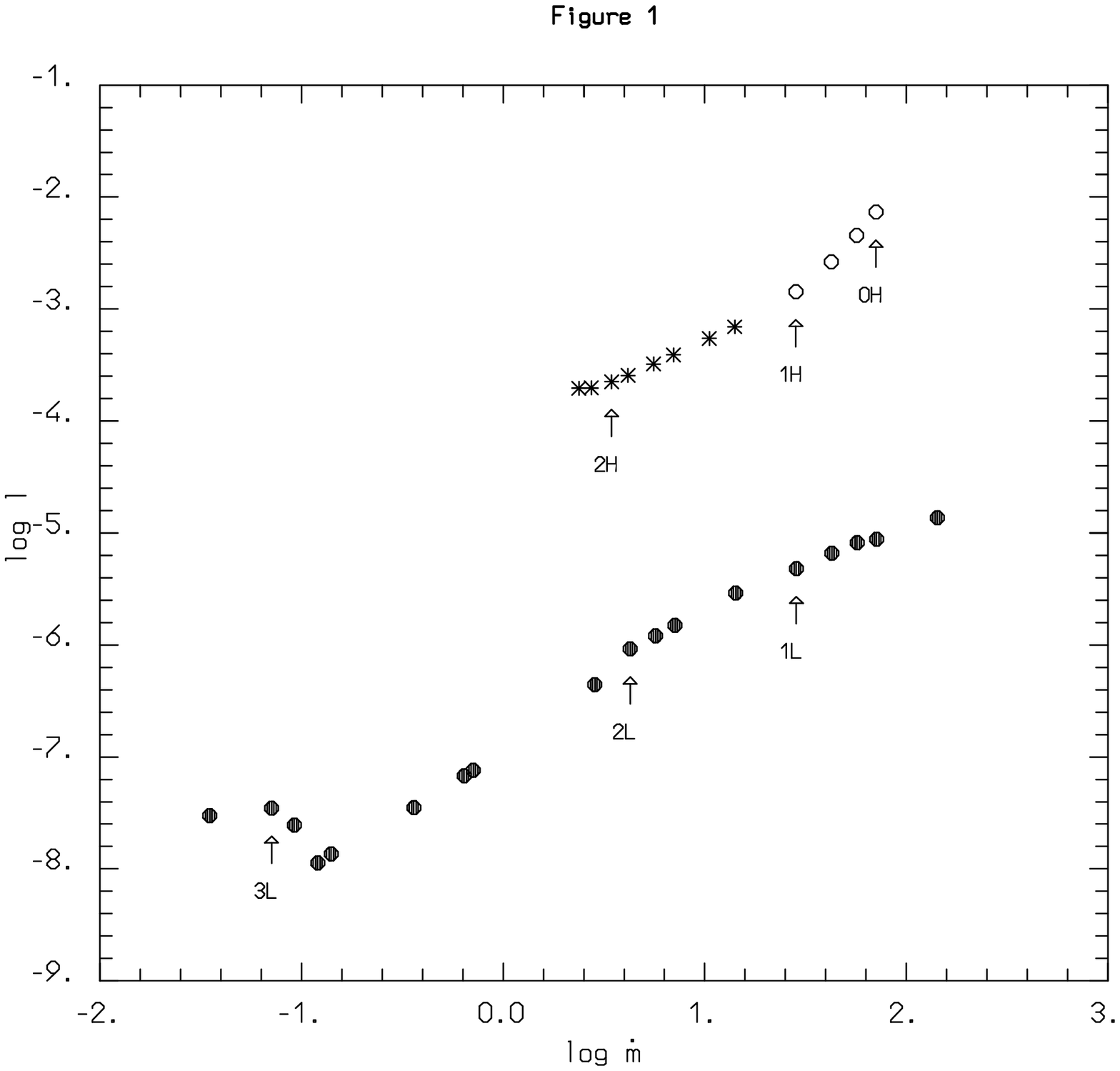,width=140mm}
\caption{{\bf Figure 1.}
 The (${\dot m}$, $l$) diagram for the
complete set of stationary solutions found by NTZ (circles). Also shown
are the 6 initial models whose relevant parameters are listed in table 1.
Filled circles mark the stable LL stationary solutions, while open circles
denote the unstable HL models. Asterisks indicate the low ${\dot m}$ HL
solutions, which might still be unstable, but on much longer time--scales.}
\endfigure
The stability of these solutions remains a completely open question
and the main goal of the present paper is to study this.
The first attempt to investigate the stability of
isothermal accretion was made by Stellingwerf \& Buff (1978)
using an Eulerian scheme, based on an extension of the Henyey relaxation
method. They found that transonic accretion is quite stable.
By means of a general relativistic analytical calculation, Moncrief (1980)
showed that, for isentropic flows, no unstable normal modes exist
which extend outside the sound horizon. The first studies including
the heating and cooling terms due to the presence of the radiation field
(Cowie et al. 1978;
Stellingwerf \& Buff 1982; Stellingwerf 1982) were devoted to analysing
the effect of preheating on the stability of the accretion flow and to defining
the region of the (${\dot m},\,  l$) plane where the existence of stationary
solutions is not allowed.
In particular, Stellingwerf (1982) presented a local stability analysis of
optically thin, X--ray heated accretion flows and showed that, for sufficiently
high luminosities, a finite amplitude drift instability can develop,
due to the form of the free--free cooling function, causing a time--dependent
behaviour of the solution on a time--scale ranging between a day and a few
tens of days.
Krolik \& London (1983) used the WKB method to derive the dispersion relation
for modes coupling
density, temperature and velocity perturbations in an optically thin,
accreting gas and found that, although stationary
solutions with high temperature and luminosity can exist, heating of the gas
inside the sonic radius leads to the onset of a thermal instability in a
large region of the (${\dot m}$, $l$) plane.
Gilden \& Wheeler (1980) and
Vitello (1984) investigated time--dependent, optically thick accretion
within the framework of General Relativity, treating the radiation field in the
diffusion approximation and using two different numerical schemes: a
Lagrangian hydrodynamic code in the first case and
a Linearized Block Implicit Algorithm in the second one.
They found that, within this approximation,
no matter which initial conditions the code started
from, convergence toward stationary LL solutions was rapidly achieved,
showing that they are intrinsically stable.
Finally, NTZ, using an argument originally suggested by
Nobili, Calvani \& Turolla (1985) and
based on Prigogine's criterion, argued that HL solutions might be
unstable because of the large value of the entropy production rate.

Despite the fact that the stationary problem has been extensively investigated,
mainly for shedding light on the efficiency of the radiation generation,
we think that several aspects require further consideration such as,
for instance, investigating the stability properties of the high luminosity
solutions and searching for the existence and non--linear evolution of possible
heated or shocked models.
In the present paper we present an analysis of the stability
and time--dependent behaviour of the solutions for spherical accretion
onto black holes within the framework of general relativistic radiation
hydrodynamics. We adopt the conventions that Greek indices
run from 0 to 3 and covariant derivatives are denoted with a semi--colon;
a spacelike signature ($-$,$+$,$+$,$+$) is used.

\section{Theory}

In this section we review the derivation of the equations of
relativistic radiation hydrodynamics in spherical symmetry
for a self--gravitating matter fluid which is interacting with radiation.
A complete treatment of this subject was recently presented by
Rezzolla \& Miller (1994) in connection with the cosmological
quark--hadron transition (see also Schmid--Burgk 1978, Mihalas \&
Mihalas 1984 and Park 1993 for a discussion of the
equations of radiation hydrodynamics in spherical flows).
Since the source terms which account for the emission and absorption of
photons are more easily written
in the reference frame comoving with the gas, this will be
our fiducial frame and $u^{\alpha}$ will denote its
4--velocity.
In this frame the energy and momentum conservation equations can be obtained by
projecting the 4--divergence of the stress--energy tensor of the matter
plus radiation fluid along $u^{\alpha}$ and orthogonal to it, giving
$$ u_{\alpha} \left( T^{\alpha\beta}_M + T^{\alpha\beta}_R
\right)_{;\beta} = 0  \, , \eqno(1)$$
$$ h_{\gamma\alpha} \left( T^{\alpha\beta}_M + T^{\alpha\beta}_R
\right)_{;\beta} = 0  \, , \eqno(2)$$
where $h_{\gamma\alpha} = g_{\gamma\alpha} + u_{\gamma} u_{\alpha}$ is the
projection tensor orthogonal to the 4--velocity and
$$ T^{\alpha\beta}_M = \left( e + p \right) u^{\alpha} u^{\beta} +
p g^{\alpha\beta} \, , \eqno(3)$$
$$ T^{\alpha\beta}_R = {\cal M} u^{\alpha} u^{\beta} +
2 {\cal M}^{( \alpha} u^{\beta )} +  {\cal M}^{\alpha \beta} +
{1\over 3} {\cal M} h^{\alpha \beta} \eqno(4)$$
are the matter and radiation stress--energy tensors, respectively;
$e$ and $p$ are the energy density and pressure of the gas.
This system of equations must be supplemented with
the rest--mass conservation equation
$$ \left(\rho u^{\alpha} \right)_{;\alpha} = 0 \, , \eqno(5) $$
where $\rho$ is the rest--mass density of the matter measured in the comoving
frame.
In equation (4) the stress--energy tensor of the radiation field is expressed
in terms of the Projected Symmetric Trace--Free (PSTF) moments (Thorne 1981)
$$ {\cal M}^{\alpha_1 ... \alpha_k} = {1\over c}
 \left( \int I n^{\alpha_1} ... n^{\alpha_k} d\Omega
 \right)^{Trace-Free} \, , \eqno(6) $$
where $I = I(x^{\alpha},p^{\alpha})$ is the specific intensity of the
radiation field and $n^{\alpha}$ is the unit vector which gives the
direction of propagation of a photon as seen in the rest frame of the
fiducial observer $u^{\alpha}$. Integration is over solid angle in the
projected space and ``Trace--Free'' denotes the consequence of the usual
tensor operation. By definition, the PSTF moments are symmetric
tensors which lie entirely in the projected space and represent the
relativistic analogue of the classical moments of the specific intensity
(see e.g. Chandrasekhar 1960).
In terms of PSTF moments, $I$ can be written as
$$ I = c \sum_{k=0}^{\infty} {{(2k+1)!!}\over {4\pi k!}}
{\cal M}^{\alpha_1 ... \alpha_k}
n_{\alpha_1} ... n_{\alpha_k} \, . \eqno(7) $$
The specific intensity obeys the general relativistic equation
of radiative transfer
$$ 2{{dN}\over {dl}} = S \, , \eqno(8) $$
where $N = (c^2/2h)(I/\nu^3)$ is the photon occupation number
($\nu = c u^{\alpha}p_{\alpha}/h$ is the photon frequency measured
in the comoving frame),
$l$ is a non--affine parameter along the photon trajectory in phase--space
and $S$ is a source function which describes the effects of
the interaction between matter and radiation, its actual form
depending on the radiative processes which are considered.
The moments of the source function, ${\cal S}^{\alpha_1 ... \alpha_k}$,
can be defined in analogy with equation (6).
If there is no interaction, $S = 0$ and $N$ is conserved along each
photon trajectory.
Moment equations can be obtained from equation (8) by inserting the
expansion (7) for $I$ (and the equivalent one for $S$)
and taking the PSTF part (i.e. projecting orthogonal to
$u^{\alpha}$, removing the trace and performing the symmetrization);
this gives rise to an infinite hierarchy of differential equations.

Finally, in order to calculate self--consistently
the metric tensor $g^{\alpha \beta}$ which describes
the geometry of the space--time,
we need to solve the Einstein Field Equations for the system
$$ R^{\alpha \beta} - {1\over 2} g^{\alpha \beta} R =
{{8 \pi G}\over c^4} \left( T^{\alpha \beta}_M + T^{\alpha \beta}_R \right)
\, . \eqno(9) $$

In spherical symmetry we can define
a local orthonormal frame comoving with the flow as
$\{{\bf e}_{\hat 0}, {\bf e}_{\hat r},
{\bf e}_{\hat \theta}, {\bf e}_{\hat \varphi}\}$, with
${\bf e}_{\hat 0} = {\bf u}$, ${\bf e}_{\hat r}$ being
in the radial direction and ${\bf e}_{\hat \theta}$,
${\bf e}_{\hat \varphi}$ being orthogonal to each other
and to ${\bf e}_{\hat r}$. Since in this case $I$ and $S$ are invariant under
rotations of the photon direction $n^{\alpha}$ about ${\bf e}_{\hat r}$,
it is possible to show that all of the components
of each PSTF moment of rank $k$ can be
evaluated in the comoving frame as functions of the radial one
$$ w_k \equiv {\cal M}^{{\hat r}...{\hat r}} = 2 \pi {{k!(2k+1)}\over
{(2k+1)!!}} {1\over c} \int I P^k(\mu) d\mu \, , \eqno(10a) $$
$$ s_k \equiv {\cal S}^{{\hat r}...{\hat r}} = 2 \pi {{k!(2k+1)}\over
{(2k+1)!!}} {h\over {c^3}}
\int \nu^3 S P^k(\mu) d\mu \, , \eqno(10b) $$
where $P^k(\mu)$ is the Legendre polynomial of order $k$, $\mu = n^{\hat r}$
and ${\hat \alpha}$ denotes the ${\bf e}_{\hat \alpha}$ component.
In particular
$$ {\cal M} = w_0 \, , \eqno(11) $$
$$ {\cal M}^{\alpha} = w_1 e_{\hat r}^{\alpha} , \eqno(12) $$
$$ {\cal M}^{\alpha \beta} = w_2 \left(
e_{\hat r}^{\alpha} e_{\hat r}^{\beta}
- {1\over 2} e_{\hat \theta}^{\alpha} e_{\hat \theta}^{\beta}
- {1\over 2} e_{\hat \varphi}^{\alpha} e_{\hat \varphi}^{\beta}
\right) \, . \eqno(13) $$
In the following, we will be interested only in
frequency--integrated moments and we will use
$w_k$ to denote the integral over $\nu$ of $w_k(\nu)$.
It is easy to see that $w_0$ is the radiation energy
density and $c w_1$ is the radial component of the radiative flux.

We next introduce the spherically symmetric, comoving--frame line element
$$ ds^2 = - a^2 c^2 dt^2 + b^2 d\mu^2 + r^2 \left( d\theta^2 +
\sin^2 \theta d\varphi^2 \right) \, , \eqno(14) $$
where $t$ and $\mu$ are the Lagrangian time and the comoving radial coordinate
(taken to be the rest mass contained within a comoving spherical shell),
$r$ is the Eulerian radial coordinate and $a$ and $b$ are two
functions of $t$ and $\mu$ which need to be computed.
The complete system of radiation hydrodynamic equations (1), (2) and (5)
along with the Einstein Field Equations (9) can be cast in the form
(Rezzolla \& Miller 1994)
$$ e_t - h \rho_t + a c s_0 = 0 \ \ \ \ \ {\rm Energy \ equation}
\, , \eqno(15) $$
$$\eqalign{
u_t & + a c \left[ {\Gamma\over b} \left( {{p_{\mu} + b s_1}\over
{\rho h}} \right) + {{4\pi G r}\over c^4} \left( p + {1\over 3}w_0 + w_2
\right) \right. \cr
& \left. + {{GM}\over {c^2 r^2}} \right] = 0 \ \ \ \ \
{\rm Euler \ equation} \, , \cr} \eqno(16) $$
$$\eqalign{
{{(\rho r^2)_t}\over {\rho r^2}} + a c & \left(
{{u_\mu - 4\pi G b r w_1/c^4}\over {r_\mu}} \right) \cr
& = 0 \ \ \ \ \ {\rm Continuity \ equation} \, , \cr} \eqno(17) $$
$$ b = {1\over {4\pi r^2 \rho}} \, , \eqno(18) $$
$$ {{(ah)_\mu}\over {ah}} + {{h\rho_\mu - e_\mu + b s_1}\over {h\rho}}
= 0 \, , \eqno(19) $$
$$ M_\mu = {{4\pi r^2 r_\mu}\over c^2} \left( e + w_0 + {u\over \Gamma}w_1
\right) \, , \eqno(20) $$
where
$$u = {{r_t}\over {ac}} \eqno(21)$$
is the radial component of the fluid 4--velocity measured in
the fixed Eulerian frame, $\Gamma = (1 + u^2 - 2GM/c^2 r)^{1/2} = r_\mu/b$ is
the  general relativistic analogue of the Lorentz factor, $M$ represents
the effective gravitational mass (for black hole + gas + radiation)
contained within radius $r$ and $h = (e + p)/\rho$ is the specific enthalpy.
Here, the subscripts $t$ and $\mu$ denote partial derivatives with respect
to the corresponding variables and $s_0$ and $s_1$ are
the radial moments of the source function $S$.
In spherical symmetry and with the line--element (14), the first
two moment equations can be written (Thorne 1981, equation [5.10c])
$$\eqalign{
{1\over {b^{4/3}r^{8/3}}}
\left( w_0 b^{4/3}r^{8/3} \right)_t
& + {c\over {abr^2}}(w_1 a^2 r^2)_\mu \cr
& + \left({{b_t}\over b} - {{r_t}\over r} \right) w_2
- a c s_0 = 0 \, , \cr} \eqno(22a) $$
$$\eqalign{
{1\over {b^2 r^2}}
\left( w_1 b^2 r^2 \right)_t
& + {c\over {3a^3 b}}(w_0 a^4)_\mu \cr
& + {c\over {b r^3}}(w_2 a r^3)_\mu
- a c s_1 = 0 \, . \cr} \eqno(22b) $$
In equations (22) $w_0$ and $w_1$ have the dimensions of
energy density and $c s_0$ and $c s_1$ are in units of erg cm$^{-3}$ s$^{-1}$.
It is well--known that the moment equations form a recursive system of
differential equations that is not closed. At any given order
$k_{max}$ it contains moments up to order $k_{max+1}$ in the
frequency--integrated case (and $k_{max+2}$ in the
frequency--dependent case). This means that, in order to use these equations
for calculations, it is
necessary to make some ``ad hoc'' assumption to close the system
(see e.g. Fu 1987, Cernohorsky \& Bludman 1994 and references therein)
and this is usually done on physical grounds by
introducing suitable closure functions which relate $w_{k_{max+1}}$
(and $w_{k_{max+2}}$ where necessary) to moments of lower order. Since the
behaviour of all of the moments is known in the asymptotic limits (when
the interaction between matter and radiation is either very strong
or completely absent), it is sufficient to
prescribe a reasonable smooth function that connects these two
limits (see e.g. NTZ). Clearly uncertainties in
this will introduce some error into the calculation of the lower
order moments, whose magnitude will be dependent on the closure relation
but turns out to be no larger than $\sim$ 15 \%  for
the range of parameter values typical for real astrophysical
flows (Turolla \& Nobili 1988).

The radiation hydrodynamic equations (15)--(21), together with
the first two moment equations (22) need to be
supplemented with the constitutive equations for the gas,
the expressions for the source moments, a prescription for the closure
function and suitable boundary conditions.

\section{The Model}

In the following, we will consider spherical accretion of a
self--gravitating hydrogen gas in the gravitational field of a
non--magnetized, non--rotating black hole.
The basic equations have been presented in the
previous section; here we will specify all of the input physics
(expressions for the source moments, equations of state, etc.)
needed for solving the problem. Stationary, spherical accretion onto
black holes has recently been investigated in detail by NTZ. The main goal
of the present paper is to ascertain the stability properties of the
solutions found by NTZ; in particular, we want to study the behaviour of the
models in a certain range of accretion rates, for which both low and high
luminosity solutions exist. To allow a direct comparison of our results with
those of NTZ, we will adopt the same input physics which they considered,
including the simplifying assumptions that Compton scattering can
always be treated in the Kompaneets limit and that pair processes
can be neglected. This turns out to be an excellent approximation for
LL models while it becomes questionable in HL solutions where the gas
temperature reaches $\sim 10^{10}$ K in the inner region. On the other
hand, the work by Park (1990b) has shown that the inclusion of pair
production and annihilation does not produce any dramatic changes.

If the dominant radiative processes are bound--free,
free--free and isotropic scattering, the radial source moments $s_0$
and $s_1$ can be written (Thorne 1981; NTZ)
$$ s_0 = \rho \left( \epsilon - k_0 w_0 \right) + k_{es}\rho w_0
{{4 k_B T}\over {m_e c^2}} \left( 1 - {T_\gamma\over T} \right) \, ,
\eqno(23) $$
$$ s_1 = - \rho k_1 w_1 \, , \eqno(24) $$
where $c \epsilon$ is the emissivity per unit mass per unit time,
$k_0$, $k_1$ and $k_{es}$
are the absorption, flux--mean and scattering opacities,
respectively, and $T_\gamma$ is the radiation temperature, defined by
$$ T_\gamma = {1\over {4k_B}} { {\int_0^\infty h \nu w_0(\nu) d\nu}
\over {\int_0^\infty w_0(\nu) d\nu} } \, . \eqno(25) $$
In equation (23) the second term on the right hand side accounts for the
energy exchange between matter and radiation due to non--conservative
scatterings and is obtained by integrating the Kompaneets equation
over frequency and neglecting the non--linear term which describes induced
emission. Since it is derived in the Fokker--Planck approximation,
this term is certainly not adequate for describing
the interaction between photons and electrons when the latter become
relativistic, as happens in some of the solutions which we have computed.
However, even in this case, the model can
give a good qualitative indication of the correct results.
We have expressed the emissivity using the interpolation of the
cooling function $\Lambda$ given by Stellingwerf \& Buff (1982)
$$ \epsilon = {{\rho \Lambda}\over {m_p^2 c}} \, , \eqno(26) $$
$$\eqalign{
\Lambda = & \left[ \left( 1.42 \times 10^{-27} T^{1/2} \beta_{rel}
+ 6.0 \times 10^{-22} T^{-1/2} \right)^{-1} \right. \cr
& \left. + 10^{25} \left( {T\over {15,849 K}} \right)^{-12}
\right]^{-1} \ \ {\rm erg \ cm}^3 \ {\rm s}^{-1} \, , \cr} $$
$$ \beta_{rel} = \left( 1 + 4.4 \times 10^{-10} T \right) \, , $$
which includes bound--bound, free--bound, $e$--$p$ and $e$--$e$
bremsstrahlung for a pure hydrogen gas; the factor
$\beta_{rel}$ is a relativistic correction. Assuming
LTE between emitters and absorbers, we can use the Kirchhoff
law to obtain the Planck mean opacity, $k_P = \epsilon / a_R T^4$,
where $a_{R}$ is the blackbody radiation constant.
Since the actual spectral distribution of $w_0$ cannot be computed here,
we use $k_P$ in place of the absorption opacity $k_0$.
The flux--mean opacity $k_1$ can be split into two terms:
the first is the scattering opacity $k_{es}$ and the second is the sum of
the contributions from all of the other radiative processes; however, since
$k_{es}$ is always dominant for the range of densities and temperatures
encountered in the present problem, we have approximated the additional term
using the Rosseland mean $k_R$ calculated taking into account only free--free
processes
$$ k_1 \simeq k_{es} + k_R \, , \eqno(27) $$
$$ k_R = 6.4 \times 10^{22} \rho T^{-7/2} \ \ {\rm cm}^2 \ {\rm g}^{-1} \, . $$
In the frequency--integrated transfer problem,
the radiation temperature $T_\gamma$ cannot be directly computed from its
definition (equation [25]).
However, since $T_\gamma$ appears only in the term in $s_0$ which accounts
for comptonization, it only becomes important when the energy exchange
between matter and radiation due to non--conservative scatterings
starts to be effective.
In a medium at rest, the fractional change of the mean
photon energy ($E = 4 k_B T_\gamma$) because of scatterings with a thermal,
non--relativistic distribution of electrons, follows the relation
$$ {{d E}\over E} = {{4k_BT}\over {m_e c^2}}
\left( {E\over {4 k_B T}} - 1 \right) \alpha d \tau \, ,$$
(Rybicki \& Lightman 1979, Wandel, Yahil \& Milgrom 1984)
where $\tau$ is the scattering depth,
$4 k_B T (E/4k_BT - 1) /m_e c^2 $ is the mean energy change per scattering
and $\alpha d \tau$ is the mean number of scatterings which a photon
undergoes between $\tau$ and $\tau + d\tau$, with
$$ \alpha = 1 \qquad \qquad \qquad \qquad {\rm for} \ \tau < 1 \, , $$
$$ \alpha = 2 \tau \qquad \qquad \qquad \qquad {\rm for} \ \tau > 1 \, . $$
From the computational point of view, it is convenient to write an
equation for $T_\gamma$ which is valid over all of the integration domain
with continuous coefficients; for this reason the
previous equation is usually written in the following form, which is
an approximation near to $\tau \simeq 1$ (Park \& Ostriker 1989; Park 1990a):
$$ {{d \ln T_\gamma}\over {d \ln \tau}} = 2 Y_c \left({T_\gamma\over T} - 1
\right) \, , \eqno(28) $$
where $Y_c = 4 k_B T \max (\tau, \tau^2)/m_e c^2$ is the Compton parameter.
In stationary calculations, this equation has been used directly to give
the variation of $\Tg$ with $r$, at constant Eulerian time $\widetilde t$,
but for non--stationary
flows it is not satisfactory to integrate it along the time--slice (i.e.
at constant Lagrangian time $t$) as this would imply an infinite speed of
propagation of information. Instead, we apply it along the outward--pointing
characteristics of the radiation field, $\mu_c (t)$ (defined from the
moment equations [22]),
and calculate the optical depth $\tau$ along the same lines using
$$ \tau = \int_\mu^\infty k_{es} \rho b \Gamma d \mu \, . \eqno(29) $$
This seems a reasonable choice because the radiation
temperature is strictly related, by definition, to the radiation energy
density and we expect that information will propagate along the
characteristic lines of the radiation field.
In this case it is not difficult to show (see the Appendix, equation [A5])
that
$$ {{\partial \Tg}\over {\partial t}} = - a c \Gamma v_c
\left[ {{2 k_{es} \rho Y_c}\over \tau} \Tg \left( {\Tg\over T} - 1 \right)
+ {1\over {b\Gamma}} {{\partial \Tg}\over {\partial \mu}} \right]
\, , \eqno(30) $$
where $v_c = b {\dot \mu}_c / a c$ is the characteristic
velocity for the radiation field and ${\dot \mu_c} = d\mu_c/dt$.
This is the actual form of the equation for $T_\gamma$ used
in our calculations. Equation (30) applies
when comptonization is the dominant radiative process and also
gives the correct behaviour (an outgoing radiation temperature wave) when
non--conservative
scattering becomes inefficient as long as true emission and absorption
can be neglected. For the HL models, true absorption is never dominant
and true emission is never likely to significantly affect $T_\gamma$
and so the use of equation (30) is always satisfactory. For the LL models,
true emission and absorption {\it are} dominant but in this case the
second term on the right hand side of equation (23) is small compared
with the other terms and an accurate evaluation of $\Tg$ is no longer needed.
The stationary limit of equation (30) (see again the Appendix,
equation [A8]) is
$$ \left( 1 + {u\over {\Gamma v_c}} \right)
{{\partial \Tg}\over {\partial r}} \barra_{\widetilde {t}} =
- {{2 k_{es} \rho Y_c}\over \tau} \Tg \left( {\Tg\over T} - 1 \right)
\, , \eqno(31) $$
where the partial derivative is taken at constant Eulerian
time $\widetilde {t}$. In this form, the presence of a critical point
where $u = - \Gamma v_c$ is made apparent. We note that this result is a
consequence of the finite velocity of propagation in equation (30);
in fact, as is well--known, the presence of critical points in the
hydrodynamic equations for stationary flow
is a relic of the characteristic velocity of
the corresponding time--dependent equations. This result represents the
main difference between the form of the $\Tg$ equation
used here and the one considered in all previous studies
of this problem in the framework of black hole accretion
(Park \& Ostriker 1989; Park 1990a; NTZ).
We stress that in all the discussion
leading to the equation for $\Tg$, we assumed a non--relativistic distribution
for electrons. This has been done to ensure consistency with the Compton
source term which has the simple expression given in equation (23) only in the
non--relativistic limit.

Finally, we need to specify the constitutive equations for a pure hydrogen gas
$$ P = [1 + x(T)] {{\rho k_B T}\over m_p} \, , \eqno(32) $$
$$\eqalign{
e = \rho c^2 \biggl\{ 1 \biggr.
& + {3\over 2} [ x(T) + x^{*}(T) ] {{k_B T}\over {m_p c^2}} \cr
& \left. - [1 - x(T)] {{E_H}\over {m_p c^2}} \right\} \, , \cr} \eqno(33) $$
where $T$ is the gas temperature, $E_H = 13.6 \ {\rm eV}$ is the hydrogen
ionization potential and $x(T)$ is the degree of ionization,
computed by equating the collisional--ionization
and radiative--recombination rates (Buff \& McCray 1974)
and expressed using the interpolation formula of Stellingwerf and Buff
(1982)
$$ x(T) = {F\over {1 + F}} \, , \eqno(34a) $$
with
$$ F = 2 \left( {T\over {1 \ {\rm K}}} \right)
\exp \left( {{-1.58 \times 10^5 \ {\rm K}}\over T} \right) \, . $$
In equation (33)
$$ x^{*}(T) = {2\over 3} \left[ \theta^{-1} \left( \eta - 1 \right)
- 1 \right] \, , \eqno(34b) $$
where $\theta = k_B T/m_e c^2$ and $\eta = K_3(\theta^{-1})
/K_2(\theta^{-1})$ ($K_n$ is the modified Bessel function of order
$n$). A polynomial fit by Service (1986) was used to calculate $\eta$,
giving an accuracy of a few parts in $10^5$.
The third term inside the curly brackets in equation (33) accounts for the
electrostatic potential energy of bound electrons in the neutral
hydrogen atoms.

The constitutive equations (32) and (33) can be used to express two
fluid variables in terms of the other ones.
Since the values of the temperature $T$ and the density $\rho$ are
needed for evaluating the source moments $s_0$ and $s_1$, it is more
convenient to use them in the hydrodynamic equations
and to calculate $e$ and $P$ from equations (32) and (33).
Substituting equation (33) into equation (15), the Energy equation can be
written in the form
$$\eqalign{
\left[ {3\over 2} \left( x + x^* \right) \right.
& \left. + \left( {3\over 2} + {{E_H}\over {k_B T}} \right)
{{dx}\over {d \ln T}}
+ {3\over 2} {{dx^*}\over {d \ln T}} \right] {{k_B T}\over {m_p c^2}}
{{T_t}\over T} \cr
& + P \left( {1\over \rho} \right)_t + {{a s_0}\over \rho} = 0
\, . \cr} \eqno(35) $$

As far as the closure function is concerned, in the present calculation
we chose to relate $w_2$ to $w_0$ using the following expression
$$ f = {w_2\over w_0} =
{2\over 3} \left[ 1 + \left({\tau\over \tau_0}\right)^n \right]^{-1} \, ,
\eqno(36) $$
where $\tau_0$ and $n$ are free parameters.
We made several trials with different expressions
for $f$ and found that the fractional difference between solutions obtained
with different reasonable closures turns out to be no larger than $\sim$
20 \%, which is acceptable here. In fact,
a change in the closure parameters was used
to perturb the initial stationary solution, as we will discuss later.
%
\begintable{1}
\nofloat
\caption{{\bf Table 1.} Parameters for the Stationary Models.}
\halign{#\hfil&\quad \hfil#\hfil\quad& \hfil#\hfil\quad& \hfil#\hfil\cr
%
%
$\,$ & ${\dot m}$ & $l$ & $\eta=l/{\dot m}$ \cr
1L & 28.5 & $4.8 \times 10^{-6}$ & $1.7 \times 10^{-7}$ \cr
2L & 4.27 & $9.2 \times 10^{-7}$ & $2.2 \times 10^{-7}$ \cr
3L & 0.071 & $3.5 \times 10^{-8}$ & $4.9 \times 10^{-7}$ \cr
0H & 70.8 & $7.3 \times 10^{-3}$ & $1.0 \times 10^{-4}$ \cr
1H & 28.3 & $1.4 \times 10^{-3}$ & $5.0 \times 10^{-5}$ \cr
2H & 3.45 & $2.2 \times 10^{-4}$ & $6.5 \times 10^{-5}$ \cr
}
\endtable

\section{Boundary Conditions}

From the mathematical point of view, the equations of radiation hydrodynamics
(16)--(21) and (35), the first two moment equations (22) and the radiation
temperature equation (30)
form a hyperbolic system. In order for the problem to be well--posed,
we need to specify values for
all of the variables at some initial time $t=t_0$ over
all of the integration domain $\mu_{in} \leq \mu \leq \mu_{out}$ and
also to assign suitable boundary conditions at the spatial boundaries
$\mu_{in}$, $\mu_{out}$.
The number of boundary conditions needed depends on the sign of the
eigenvalues of the characteristic equations at each boundary. At the
outer boundary, negative eigenvalues signify that information
is propagating into the integration domain from outside and a corresponding
number of conditions must be assigned; the same is true for positive
eigenvalues at the inner boundary. In the present case it can be shown that we
need to prescribe 7 boundary conditions (4 at the inner boundary and 3 at the
outer boundary) as follows: 2 conditions related to the fluid equations,
2 related to the moment equations, 1 each for the equations for $\Tg$, $a$
and $M$. As far as the
fluid boundary conditions are concerned: at $\mu = \mu_{out}$
we set a floating boundary (extrapolation in $r$) for $u$ and, at
$\mu = \mu_{in}$,
we dropped the pressure gradient term from the Euler equation,
making it advective in form and assuming strict free--fall very near to
the black hole horizon.
The inner conditions on $\Tg$ and $w_1$ and the outer condition on $w_0$
were all taken as floating boundaries. The choice of a floating
boundary is suitable when one does not want to put any constraint on a
variable, leaving it free to adjust itself or to oscillate if there are
waves propagating out of the integration domain (as for a vibrating
string with free endpoints).

As far as $a$ and $M$ are concerned, the time--slice at
constant $t$ is a characteristic direction for equations (19) and (20)
and we put
$$ a = 1 \ \ \ \ \ \ \ \ \ \ \ \ \ \ \
{\rm at} \ \mu = \mu_{out} \, , \eqno(37) $$
$$ M = M_0 \ \ \ \ \ \ \ \ \ \ \ {\rm at} \ \mu = \mu_{in} \, . \eqno(38) $$
The condition on $a$, equation (37), corresponds to synchronizing the
coordinate time with the proper time of a comoving observer at the outer
edge of the grid. This is also equal to the Eulerian time there,
if the outer edge of the grid is placed sufficiently far away from the
black hole.

Finally, we note that, if the system tends to a stationary limit,
the time--dependent equations reduce to their stationary form and
the solution which crosses any critical points in a regular way is
automatically selected.
%
\begintable*{2}
\caption{{\bf Table 2.} Time--scales.}
\halign{#\hfil&\quad \hfil#\hfil\quad& \hfil#\hfil\cr
%
%
$ \left. t_d = {r\over {uc}}\right. $ & dynamical time & $\,$ \cr
$ \left. t_{sg} = {r\over {v_s c}} \right. $
& sound crossing time for the gas &
$v_s = {1\over c} \left( {{\partial P}\over {\partial \rho}} \right)_s^{1/2} $
\cr
$ \left. t_{sr} = {r\over {v_c c}} \right. $
& sound crossing time for the radiation &
$ v_c = \left( f + {1\over 3} \right)^{1/2} $ \cr
$ \left. t_c = {{e-\rho c^2 + P}\over C} \right. $
& cooling time &
$ C = \rho c \left( \epsilon + k_{es} w_0 {{4k_BT}\over {m_e c^2}}
\right) $ \cr
$ \left. t_h = {{e-\rho c^2 + P}\over H} \right. $
& heating time &
$ H = \rho c w_0 \left( k_0 + k_{es} {{4k_BT_{\gamma}}\over
{m_e c^2}} \right) $ \cr
$ \left. t_{th} = {{e-\rho c^2 + P}\over {\vert H - C
\vert}} \right. $
& thermal time & $\,$ \cr
}
\endtable

\section{Numerical Method}

The equations of radiation hydrodynamics, presented
in sections 2 and 3, have been solved numerically
for matter being accreted spherically onto a
Schwarzschild black hole, using a Lagrangian finite difference scheme
with a standard Lagrangian organization of the grid.
The code was adapted from one developed by Miller \& Rezzolla (1995)
for solution of the equations of radiation hydrodynamics in the context of
the cosmological quark--hadron transition.
We divided the integration domain (from the black hole horizon at $r=r_g$
out to $10^9 r_g$) into a succession of comoving zones with each one having
width $\Delta \mu$ 21\% larger than the one interior to it.
Whenever the inner edge of the innermost zone crosses the horizon
(which happens every 4--5 cycles with our time--step constraints),
we remove it from the calculation and perform a regridding
of all the variables. We followed a regridding procedure previously adopted
by Szuszkiewicz \& Miller (1995) in connection with the
study of disc accretion onto black holes. Originally
a cubic spline interpolation was used, but this turned out to produce
a numerically unstable evolution in our case.
A local cubic interpolation was eventually used instead, which was found to be
satisfactory and efficient.
At the initial time the effective mass contained within the inner boundary,
$M_0$, is equal to the initial black hole mass $M_{bh}$, which we take to be
$3M_\odot$ as in the stationary calculations. As time elapses, $M_0$ increases
as zones pass through the horizon and are removed from the calculation.
However, during a characteristic evolutionary time interval,
the mass of the material accreted is small compared with $M_{bh}$.

To have second--order accuracy in time, $u$ and $w_1$ are both evaluated
at an intermediate time level.
They are advanced to the new time level at the end of each cycle after
all of the other variables have been
calculated. The time--step is adjusted in accordance with
the relativistic Courant condition and two additional
constraints on the fractional variations of $\rho$ and $T$
in each time--step, which are required to be smaller than 5\%. In practice,
we found that the time--step is usually limited by the last two
conditions due to the fact that the variation of
density and temperature, as seen by comoving observers, becomes very
rapid near to the horizon where the flow velocity approaches
the speed of light. As far as the spatial centering is concerned,
$\rho$, $T$, $w_0$, $\Tg$ and $a$ are treated as mid--zone quantities,
while $r$, $M$, $u$ and $w_1$ are treated as zone boundary quantities.

Once the finite difference representation has been introduced, equations (16),
(17), (19), (20) and (21) can be solved explicitly for $u$, $\rho$, $a$,
$M$ and $r$, respectively. Where necessary, linear interpolation and
extrapolation in time were used to obtain the values of quantities at
suitable time levels.
The semi--logarithmic derivatives present in the Continuity equation
(17) and the equation for $a$ (19) were solved using the
Crank--Nicholson operator for equation (17) and the Leith--Hardy operator for
equation (19) (see e.g. May \& White 1967).
For the moment equations (22), we adopted
a mixed representation: after dividing the first equation by $w_0$ and
the second by $w_1$ we grouped together the terms in the following way:
$$ \eqalign{
{{(w_0)_t}\over w_0} = & - \left[ {4\over 3} \left( {b_t\over b}
+ 2{r_t\over r} \right) + \left( {b_t\over b} - {r_t\over r} \right) f
\right] \cr
& + {{ac}\over w_0} \left[ s_0 - {1\over {a^2 b r^2}} (w_1 a^2 r^2)_\mu
\right] \, , \cr} \eqno(39a) $$
$$ \eqalign{
{{(w_1)_t}\over w_1} = & - 2 \left( {b_t\over b} + {r_t\over r} \right) \cr
& + {{ac}\over w_1} \left[ s_1 - {1\over {3 a^4 b}} (w_0 a^4)_\mu
- {1\over {a b r^3}} (f w_0 a r^3)_\mu
\right] \, , \cr} \eqno(39b) $$
where $w_2$ has been expressed using the closure relation $w_2 = f w_0$
with $f$ being defined as in equation (36).
The terms on the left hand side of equations (39) were treated
using the Crank--Nicholson operator, while the quantities appearing on the
right hand side were calculated at the correct time level by means of
interpolation or extrapolation where necessary.
Because the dependence on temperature in the Energy equation is
rather sensitive, we adopted a semi--implicit scheme for equation (35) using
a secant iteration method. The temperature at the new time level is calculated
iteratively starting from two initial estimates, based on the value
at the previous time--step. Convergence is rapidly
achieved in 4--5 iterations. Since $s_0$ is in general very sensitive to the
value of the temperature, the iteration was extended to include also the
zero--th moment equation (39a).
The equation for $\Tg$ (equation [30]) presents a particular numerical problem
because of the delicate balance between $T$ and $\Tg$ which,
following Bowers \& Wilson (1991), we treated using a fully implicit
differencing for $(T-\Tg)$ to achieve numerical stability.

\section{Numerical Results and Discussion}

We have calculated the time evolution of 6 models, starting from the stationary
solutions listed in table 1 whose position
in the (${\dot m}$, $l$) plane is shown in Fig.~1. Because of
the different form of the $\Tg$ equation used here, the present stationary
solutions differ slightly from those of NTZ (see equation [31] and the
following discussion).
The definitions of relevant time--scales for the present discussion are
listed in table 2.
Along with the dynamical time--scale, $t_d$ (which is the
characteristic time for an element of fluid with velocity $u$ to travel a
distance $r$), we have listed also the sound crossing times for the
gas, $t_{sg}$, and for the radiation, $t_{sr}$ (these are defined as the ratio
of the radial length scale to the relevant characteristic velocity and
represent the time needed for a ``sound'' wave in the gas or
radiation fluid to travel a distance $r$). For determining whether stationary
solutions are stable or not,
we should in principle evolve each model for a time comparable with the
relevant $t_{sg}$ or $t_{sr}$ in order to allow information to travel
from the inner regions to the outer ones. At $10^5 r_g$,
$t_{sg} \sim 10^4$ s and $t_{sr} \sim 1 $ s. To reach an evolutionary time
$t = 10^4$ s would require prohibitively high computational times but
in fact, as we shall see, all of the models evolve on a much shorter
time--scale (typically of a few seconds) which is mainly determined
by the thermal and radiative processes.
The thermal balance is regulated by the cooling and heating
time--scales, $t_c$ and $t_h$, which are the ratios of the internal
energy of the gas to the cooling rate ($C$) and heating rate ($H$),
respectively, and are both defined in table 2. We have also introduced
the thermal time--scale, $t_{th}$, defined as the ratio of the
internal energy of the gas to the net rate of energy input or output
for the gas, $\vert H - C \vert$.

\beginfigure*{2}
\hskip 0.3 truecm
\psfig{figure=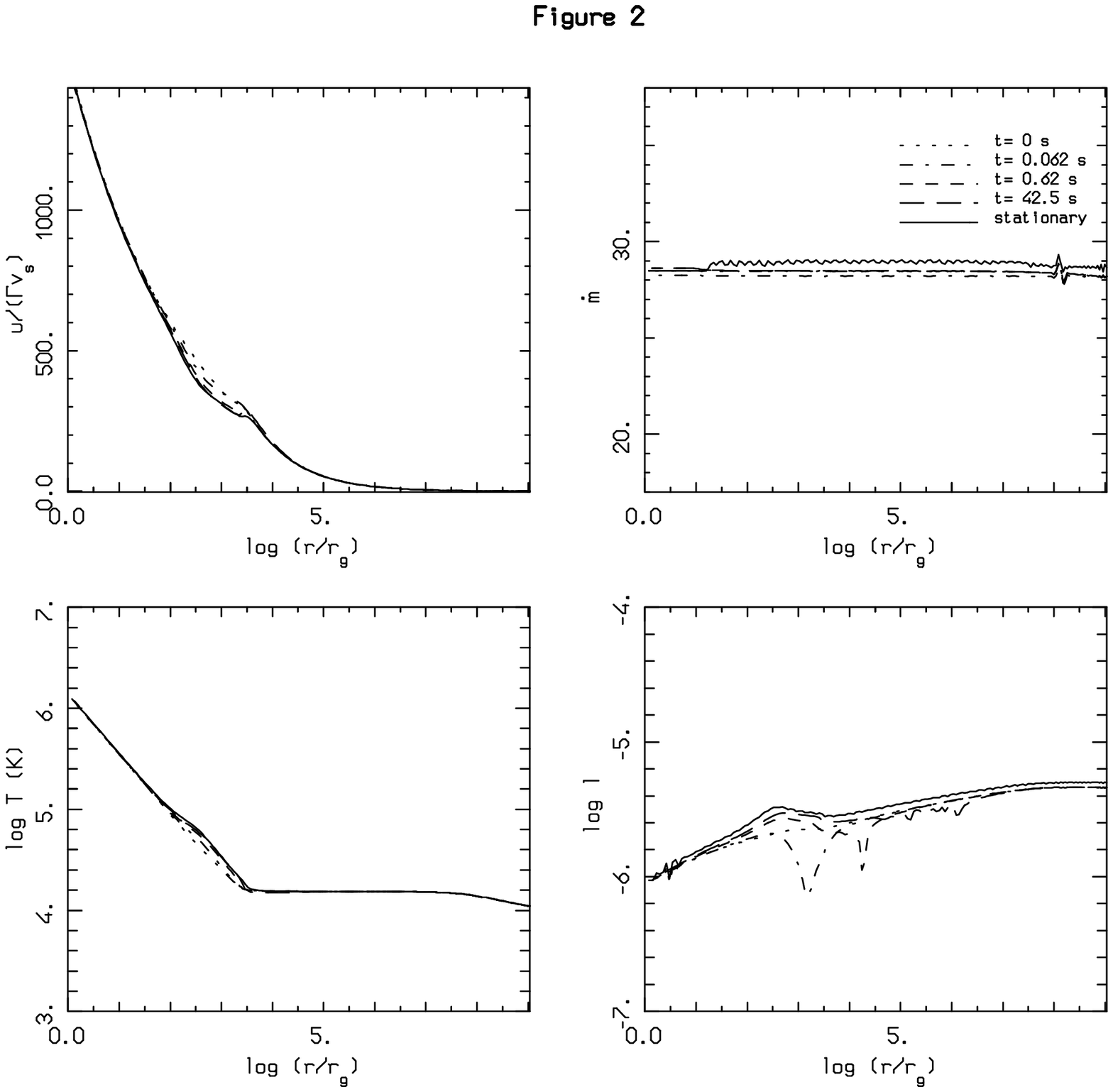,width=160mm}
\caption{{\bf Figure 2.}
The Mach number $u/\Gamma v_s$,
the accretion rate in Eddington units ${\dot m}$, the gas temperature
$T$ and the radiative luminosity $l = 4\pi r^2 c w_1/L_E$
are plotted versus $\log (r/r_g)$ for model 1L at different
times.}
\endfigure

For each model
we started the numerical calculation from an initial perturbed solution
which was calculated by changing the closure function. By varying $\tau_0$
and $n$ in equation (36), we obtained
a perturbation of the order of 10--20 \% in the gas temperature and
in the radiation moments. This way of setting the perturbation was not effective
for model 3L, which is optically thin everywhere, and so, only in this case,
we decided to evolve the solution without any initial perturbation just to have
an indication of the intrinsic stability of these models. The solution
after 14 seconds remains exactly in the initial stationary state.
This is not surprising since, in this case, we know that cooling is very
inefficient and so the result obtained by Moncrief (1980), who found that
adiabatic flows are stable, was likely to apply here.
In fact, optically thin
solutions are not of great interest and we did not spend further time in the
numerical analysis of the stability of these models.

In Fig.~2 the results from the numerical calculation
for model 1L are shown. This solution is representative
of the behaviour of all optically thick LL models.
As can be seen from the figure,
the solution relaxes toward the stationary state (shown with a continuous line)
on a time--scale of the order of $t_{th}$
which, for this solution, is much shorter than 1 second within $10^3 r_g$.
This shows that these solutions are stable, in
agreement with the result obtained by Vitello (1984).
The perturbation does not directly involve velocity and density, which
remain essentially equal to their initial values and the accretion rate also
remains extremely constant.
Radiation and gas pressure have negligible effect in these cases
and matter is essentially in free--fall from the sonic radius
(located at $r \simeq 10^9 r_g$) down to the black hole horizon.
Temperature and luminosity relax very quickly to their stationary values. In the
optically thick inner core, matter and radiation are in LTE and the luminosity
is proportional to the local value of the temperature gradient; in the outer
region compressional heating balances free--bound cooling and the gas is
essentially in radiative energy equilibrium at the hydrogen recombination
temperature, $T \simeq 10^4 K$.

\beginfigure*{3}
\hskip 0.3 truecm
\psfig{figure=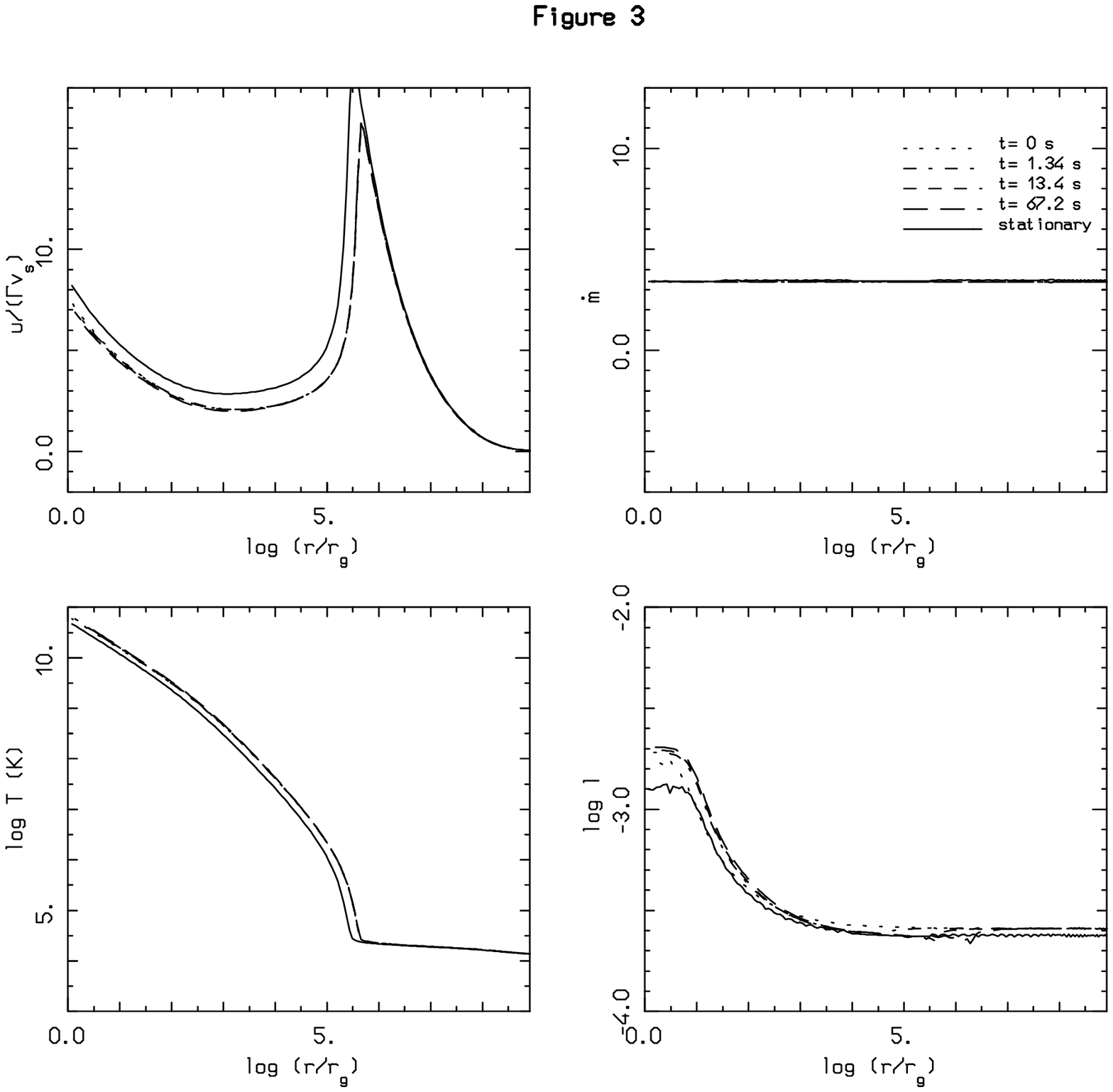,width=160mm}
\caption{{\bf Figure 3.}
The same as Fig.~2 for model 2H.}
\endfigure
\beginfigure*{4}
\hskip 0.3 truecm
\psfig{figure=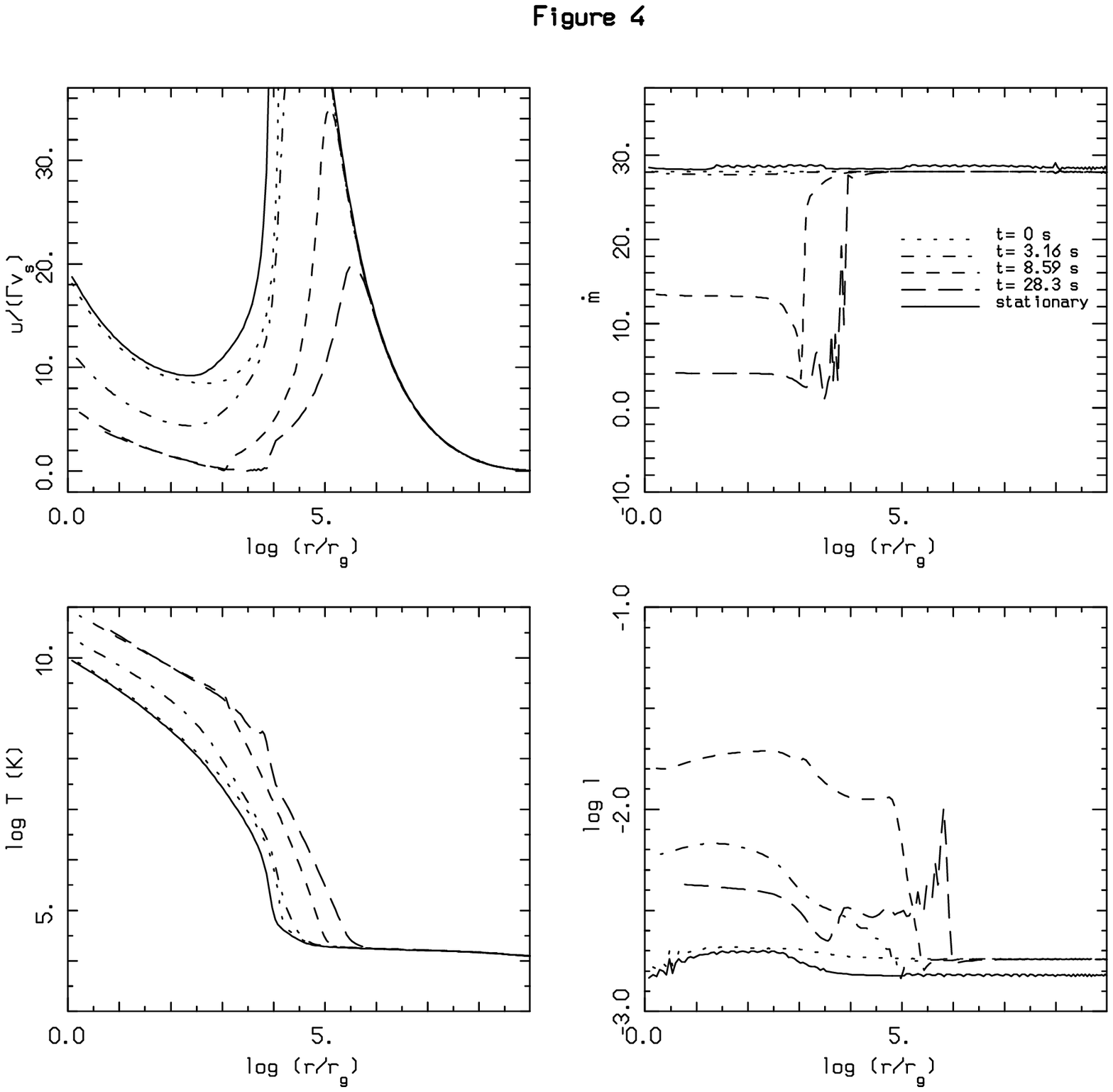,width=160mm}
\caption{{\bf Figure 4.}
The same as Fig.~2 for model 1H.}
\endfigure
For models on the high luminosity branch, the behaviour of the
time--dependent solutions is completely different. In Figs.~3 and 4,
we show the results of numerical calculations for models 2H and
1H, respectively.
The matter fluid is essentially
in free--fall and the accretion rate is roughly constant.
The interpretation of the temperature and luminosity profiles is less
straightforward. They do not seem
to converge at all toward the stationary solutions found by NTZ,
but, after 70 seconds, they are still in essentially the initial
perturbed state. We postpone
discussion of these solutions to the end of this section.

For larger $\dot m$ (model 1H),
a thermal instability appears in the inner part of the flow
after about 2--3 seconds, as can be seen from Fig.~4, and the
temperature increases by almost an order of magnitude.
The cause of the onset of this instability will be discussed later.
A few seconds after this,
the velocity profile starts to deviate significantly from free--fall
owing to the large drag exerted by the internal pressure gradients.
A compression wave develops, whose front becomes progressively steeper
as it propagates outward and, after 8--10 seconds, a hydrodynamic
shock forms at around $10^3$--$10^4 r_g$, where the Mach number passes
through unity (small--dashed line in Fig.~4).
Across the shock, the kinetic energy of the gas is dissipated into thermal
energy and the density increases; matter starts to accumulate at the shock
front and a corresponding decrease in $\rho$ is seen in the inner region.
Immediately behind the shock front, the
gas accelerates and free--fall is rapidly restored.
We note that since the shock is very far away from the black hole horizon,
the kinetic energy dissipated there is relatively small and the radiative
luminosity does not increase significantly through it.
As far as we know, this is the first time that a shock has been found in
self--consistent solutions of black hole accretion (see Chang \& Ostriker
1985 for a discussion of shock formation in stationary models).
The large increase of $l$ in the first 8--10 seconds (by more than one order
of magnitude) is due to the enhancement in efficiency of free--free and
Compton cooling caused by the increase in $T$.
At later times, $l$ starts to decrease because of the fall in density
which leads to a decrease in efficiency of
the cooling processes interior to the shock radius.
The luminosity profile then has the typical behaviour shown in Fig.~4
with the long--dashed line.
This time variation is seen by a distant observer as
a significant initial transient increase lasting
$\sim$ 8 s followed by a relatively rapid decrease at later times.
Looking carefully at the Mach number profile, it can be seen that the shock
front (where the Mach number falls below one)
is moving outward, at an approximate speed of $10^8$ cm/s $\simeq
10^{-2} c$. Hence this solution is definitely non--stationary as
confirmed by the fact that the accretion rate is not constant and matter
keeps accumulating at the shock front.
The evolution of model 1H (Fig.~4) was followed
inserting a source of artificial viscosity, with the aim of getting a better
treatment of the shock region.
Following the standard
prescription by von Neumann and Richtmyer (1950), a dissipative
term $Q \propto \rho_{i-1/2} \left(u_i - u_{i-1} \right)^2$,
was inserted into the equations (here the subscripts indicate the locations
on the finite difference grid at
which each variable is evaluated; $i$ represents a zone boundary and $i-1/2$
represents a mid--zone).
However, since the flow is being compressed continuously
from the sonic radius down to the black hole horizon, the amount of
dissipation would be excessive, especially in the vicinity of $r_g$,
unless some modification is made to the standard procedure.
In view of this, we decided to switch on the artificial
viscosity only when the fractional variation of $u$ across a grid zone
$\alpha = 2(u_i - u_{i-1})/(u_i + u_{i-1})$ becomes
larger than 30\%. As the shock forms, $\alpha$ increases above 0.3 and
the viscous term
$$ Q_{i-1/2} = k \left( 1 - {{0.3}\over \alpha} \right)
\rho_{i-1/2} \left(u_i - u_{i-1} \right)^2 c^2 \eqno(40) $$
then starts to be progressively more effective. Here $k$ is an
adjustable constant
($k=2$ in the actual calculation).
About 30 seconds after the beginning of the evolution and approximately
20 seconds after the shock formation, we were nevertheless forced to stop the
calculation because of the formation of large numerical oscillations at the
shock front
and some more sophisticated treatment would clearly be desirable.
However, for the present purposes, we were content simply to demonstrate
the existence of the shocked solutions.

The evolution described for model 1H, with the appearance of a thermal
instability and the formation of a hydrodynamic moving shock, is a common
feature of all of the high luminosity models along the high $\dot m$ part of the
HL branch. We have made a systematic search for the point on the (${\dot m}$,
$l$) plane which marks the onset of the instability and the result is shown
in Fig.~1, where all of the HL unstable models are plotted as open
circles. According to the analysis of
Field (1965) and Stellingwerf (1982), the form of the free--free cooling
function implies that the gas should be thermally unstable to isobaric
short--wavelength perturbations so that, if the Compton heating rate
exceeds the cooling rate at some radius, it will continue to do so until matter
there has been heated to a temperature which is essentially equal to that of
the radiation. In the present case, owing to the large
value of the Compton parameter $Y_c$, Compton cooling is equally as efficient
as Compton heating and the analysis by Stellingwerf (1982) does not strictly
apply.
However, as discussed by Cowie, Ostriker \& Stark (1978),
the instability is clearly due to the fact that
the heating rate is greater than the cooling rate and, at the same time,
the heating time is shorter than the dynamical time.
In Figs.~5 and 6, we have plotted the ratios of the heating time ($t_h$)
to the cooling time ($t_c$) and of the thermal time ($t_{th}$) to the
dynamical time ($t_d$). These quantities are plotted against $r/r_g$
at different times for models 2H and 1H, respectively.
\beginfigure*{5}
\hskip 2.5 truecm
\psfig{figure=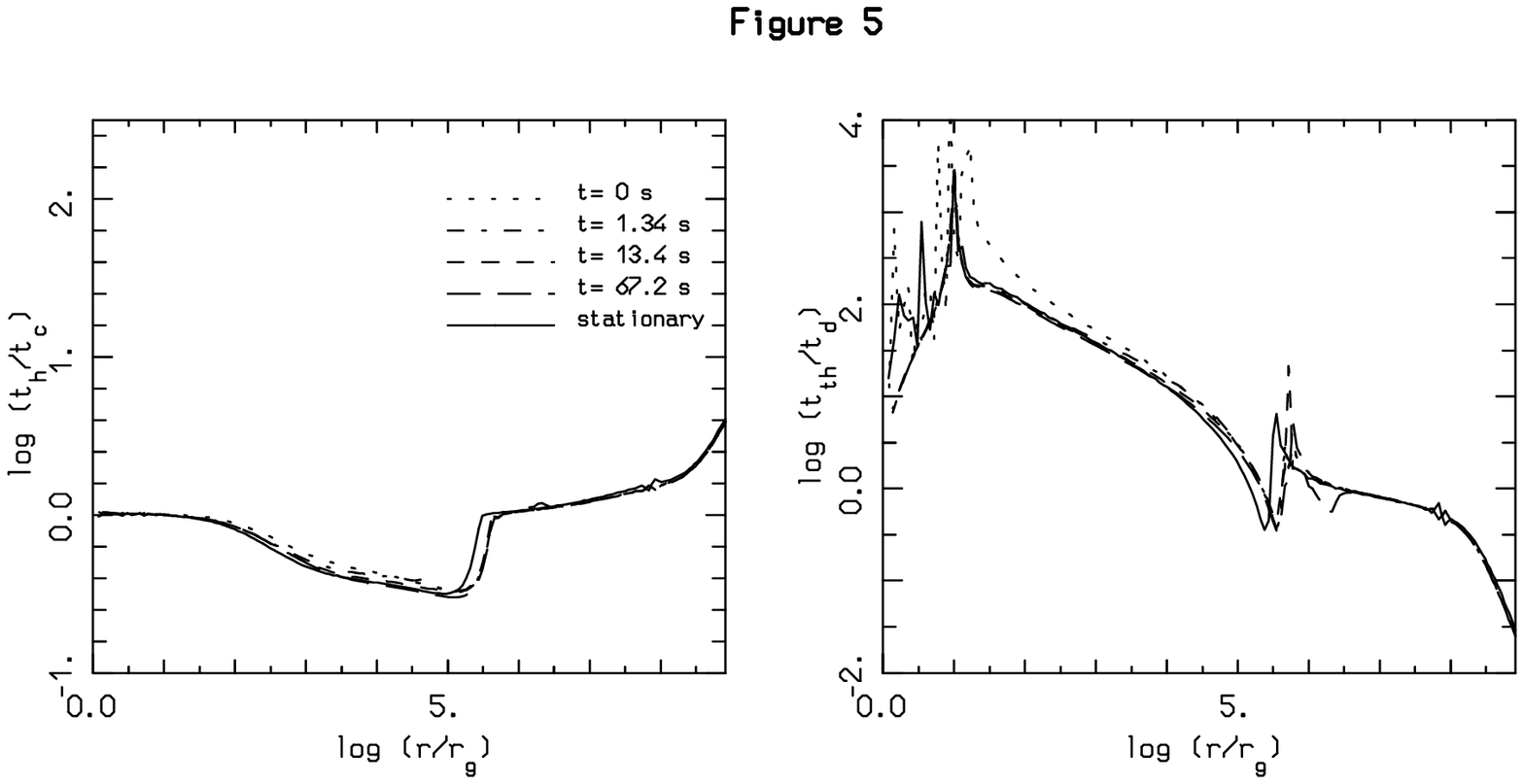,width=120mm}
\caption{{\bf Figure 5.}
Ratios of the heating time ($t_h$)
to the cooling time ($t_c$) and of the thermal time ($t_{th}$) to the
dynamical time ($t_d$) plotted against $\log (r/r_g)$ for model 2H at different
times.}
\endfigure
\beginfigure*{6}
\hskip 2.5 truecm
\psfig{figure=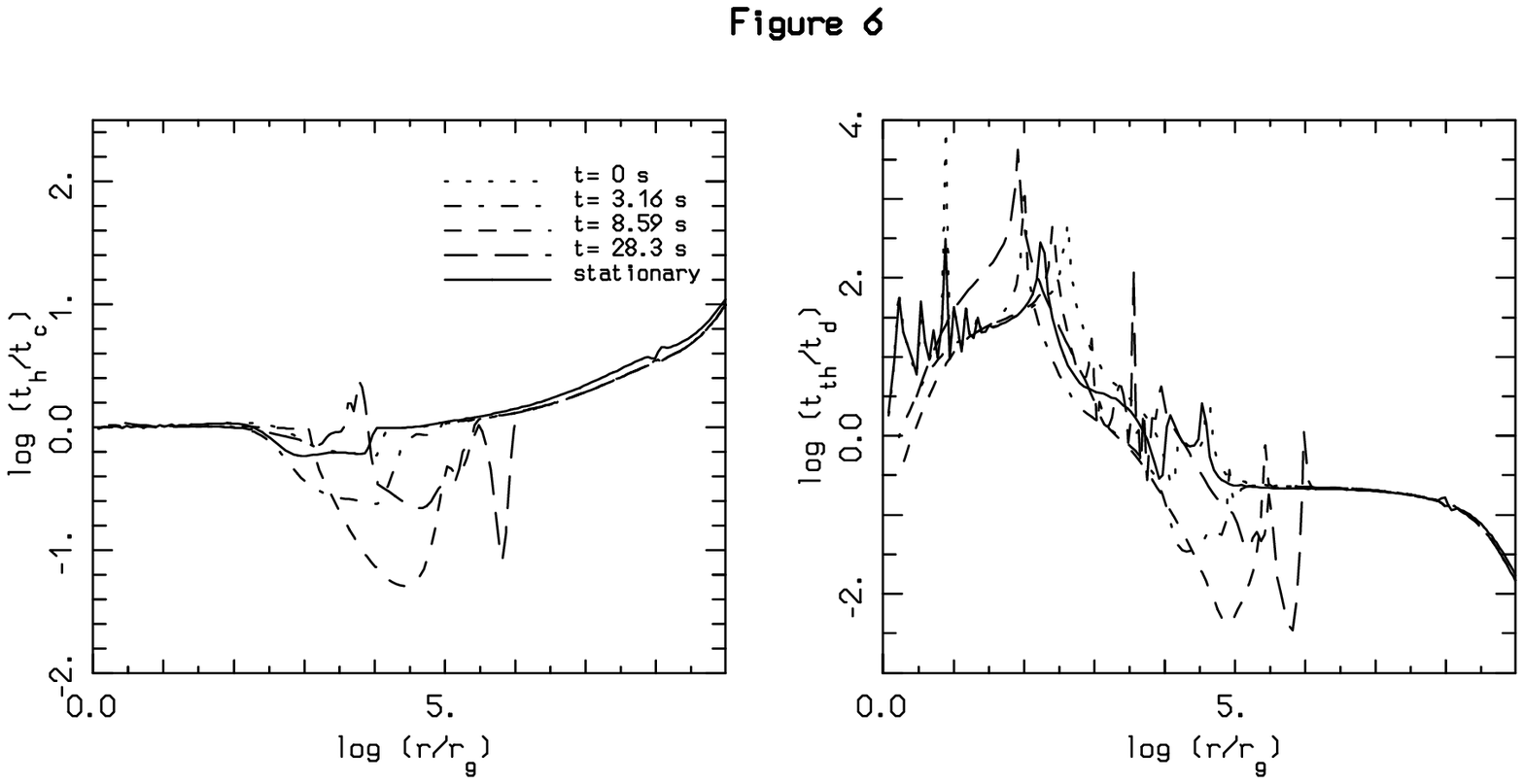,width=120mm}
\caption{{\bf Figure 6.}
The same as Fig.~5 for model 1H.
Note that the curve for $t=0$ in the left panel is difficult to see
because it is almost completely hidden by the continuous line.}

\endfigure
As can be seen from Fig.~6, at the beginning $t_h$ is slightly
smaller than $t_c$
in the region around $10^3$--$10^4 r_g$.
There, heating exceeds cooling and, since the
flow cannot advect the excess energy efficiently ($t_d \simeq t_{th}$),
a small perturbation is sufficient to make heating more effective and
the onset of a thermal instability is unavoidable.
Also the value of the thermal time--scale, $t_{th} (10^3 r_g) \sim
t_d (10^3 r_g) \sim 1$ s, is consistent with the time--scale
for the onset of the instability found numerically.
The region of instability then moves to larger radii, as can be seen from
Fig.~6, and so it is not surprising that the shock keeps moving outward.
On the other hand, Fig.~5 shows that the ratio $t_{th}/t_d$ is significantly
larger than unity for model 2H in the region
where heating is more effective than cooling and so the thermal instability is
advected into the hole on a dynamical time--scale.
In other words, since the models along the low ${\dot m}$
part of the HL branch have smaller gas densities, the radiative heating and
cooling are comparatively less efficient than compressional heating and the
gas is essentially adiabatic. Only very far away, around $3 \times 10^5 r_g$,
do the conditions seem to be favourable for the onset
of the thermal instability; however, the thermal time--scale in that region is
$\sim 10^3$ s and the evolutionary time becomes very large. This means that
these solutions might also be unstable but on much longer time--scales.

Finally, we note that, during the evolutionary times considered, we did not
find any evidence for possible transitions between the HL and the LL branch.

\section{Conclusions}

We have presented a systematic analysis of the stability and time--dependent
behaviour of spherical accretion onto black holes in the framework of
general--relativistic radiation hydrodynamics.
We have computed the evolution of a number of models from an intial perturbed
state and confirmed that all of the low luminosity, NTZ
stationary solutions, which are characterized by negligible comptonization,
are indeed stable to thermal and radiative perturbations in agreement with
previous investigations (e.g. Gilden \& Wheeler 1980;
Vitello 1984). On the other hand, the time evolution of high luminosity
solutions,
for which self--comptonization of
bremsstrahlung photons is the main radiative and
thermal process, exhibits a much richer phenomenology.
We find that the upper part of the HL branch (for ${\dot m} \ga 10$)
enters the region of the (${\dot m}$, $l$) plane where preheating effects start
to be important and this leads to the onset of a strong thermal instability
giving rise to the formation of an outward--propagating hydrodynamic
shock. These shocked solutions show significant transient
increases in the total luminosity.

\section*{Acknowledgements}

We gratefully acknowledge helpful discussions with Luciano Rezzolla,
Alan Ball and Henk Spruit.

\section*{References}

\beginrefs

\ref{Begelman, M.C. 1978, A\&A, 70, 583}
\ref{Begelman, M.C. 1979, MNRAS, 187, 237}
\ref{Bisnovatyi--Kogan, G.S., \& Blinnikov, S.I. 1980, MNRAS, 191, 711}
\ref{Blumenthal, G.R., \& Mathews, W.G. 1976, ApJ, 203, 714}
\ref{Blondin, J.M. 1986, ApJ, 308, 755}
\ref{Bondi, H. 1952,  MNRAS, 112, 195}
\ref{Bowers R.L., \& Wilson J.R. 1991, {\it Numerical Modeling in Applied
Physics and Astrophysics}, (Boston: Jones \& Bartlett Publishers)}
\ref{Brinkmann, W. 1980, A\&A, 85, 146}
\ref{Buff, J., \& McCray, R. 1974, ApJ, 189, 147}
\ref{Cernohorsky, J., \& Bludman, S.A. 1994, ApJ, 433, 250}
\ref{Chandrasekhar, S. 1960, {\it Radiative Transfer}, (New York: Dover)}
\ref{Chang, K.M., \& Ostriker, J.P. 1985, ApJ, 288, 428}
\ref{Cowie, L.L., Ostriker, J.P., \& Stark, A.A. 1978, ApJ, 226, 1041}
\ref{Field, G.B. 1965, ApJ, 142, 431}
\ref{Flammang, R.A. 1982, MNRAS, 199, 833}
\ref{Flammang, R.A. 1984, MNRAS, 206, 589}
\ref{Freihoffer, D. 1981, A\&A, 100, 178}
\ref{Fu, A. 1987, ApJ, 323, 227}
\ref{Gilden, D.L., \& Wheeler, J.C. 1980, ApJ, 239, 705}
\ref{Gillman, A.W., \& Stellingwerf, R.F. 1980, ApJ, 240, 235}
\ref{Kafka, P., \& M\`esz\`aros, P. 1976,  Gen. Rel. Grav.,  7, 841}
\ref{Krolik, J.H. \& London, R.A. 1983, ApJ, 267, 18}
\ref{Maraschi, L., Reina, C., \& Treves, A. 1974, A\&A, 35, 389}
\ref{May, M.M., \& White, R.H. 1967, {\it Methods in Computational
Physics, Vol. 7}, Alder, B., Fernbach, S. \& Rotenberg, M. eds.,
(New York: Academic Press)}
\ref{M\`esz\`aros, P. 1975, A\&A, 44, 59}
\ref{Michel, F.C. 1972, Ap\&SS, 15, 153}
\ref{Mihalas, D., \& Weibel Mihalas, B. 1984, {\it Foundations of
Radiation Hydrodynamics} (Oxford: Oxford University Press)}
\ref{Miller, J.C., \& Rezzolla, L. 1995, Phys. Rev. D, 51, 4017}
\ref{Moncrief, V. 1980, ApJ, 235, 1083}
\ref{Nobili, L., Calvani, M., \& Turolla, R. 1985, MNRAS, 214, 161}
\ref{Nobili, L., Turolla, R., \& Zampieri, L. 1991, ApJ, 383, 250, (NTZ)}
\ref{Novikov, I.D., \& Thorne, K.S. 1973, in {\it Black Holes}
DeWitt, C., \& DeWitt, B.S. eds., p. 343, (New York: Gordon \& Breach)}
\ref{Ostriker, J.P., McCray, R., Weaver, R., \& Yahil, A. 1976, ApJ, 208, L61}
\ref{Park, M.--G., \& Ostriker, J.P. 1989, ApJ, 347, 679}
\ref{Park, M.--G. 1990a, ApJ, 354, 64}
\ref{Park, M.--G. 1990b, ApJ, 354, 83}
\ref{Park, M.--G. 1993, A\&A, 274, 642}
\ref{Rees, M.J. 1978, Phys. Scripta, 17, 193}
\ref{Rezzolla, L., \& Miller, J.C. 1994, Class. Quantum Grav., 11, 1815}
\ref{Rybicki, G.B., \& Lightman, A.P. 1979, {\it Radiative Processes in
Astrophysics\/} (New York: Wiley)}
\ref{Schmid--Burgk, J. 1978, Ap\&SS, 56, 191}
\ref{Service, A.T. 1986, ApJ, 307, 60}
\ref{Shapiro, S.L. 1973a, ApJ, 180, 531}
\ref{Shapiro, S.L. 1973b, ApJ, 185, 69}
\ref{Shull, J.M. 1979, ApJ, 229, 1092}
\ref{Shvartsman, V.F. 1971, Soviet Astr.--AJ, 15, 377}
\ref{Soffel, M.H. 1982, A\&A, 116, 111}
\ref{Stellingwerf, R.F., \& Buff, J. 1978, ApJ, 221, 661}
\ref{Stellingwerf, R.F., \& Buff, J. 1982, ApJ, 260, 755}
\ref{Stellingwerf, R.F. 1982, ApJ, 260, 768}
\ref{Szuszkiewicz, E., \& Miller J.C. 1995, in preparation}
\ref{Tamazawa, S., Toyama, K., Kaneko, N., \& ${\hat {\rm O}}$no, Y.
1974, Ap\&SS, 32, 403}
\ref{Thorne, K.S. 1981, MNRAS, 194, 439}
\ref{Turolla, R., \& Nobili, L. 1988, MNRAS, 235, 1273}
\ref{Vitello, P.A.J. 1978, ApJ, 225, 694}
\ref{Vitello, P.A.J. 1984, ApJ, 284, 394}
\ref{von Neumann, J., \& Richtmyer, R.D. 1950, J. Appl. Phys., 21, 232}
\ref{Wandel, A., Yahil, A., \& Milgrom, M. 1984, ApJ, 282, 53}

\endrefs

\section*{Appendix A}


As discussed in the text, we take equation (28) to be valid along the
outward--pointing characteristics of the radiation field, $\mu_c (t)$,
with the optical depth being defined along the same lines by
$$ \tau = \int_\mu^\infty k_{es} \rho b \Gamma d \mu \, . \eqno({\rm A}1) $$
Then, we have
$${{d \tau}\over {d \mu}} = - k_{es} \rho b \Gamma \, . \eqno({\rm A}2) $$
(Throughout this Appendix the total derivatives are taken along the
outward--pointing characteristics of the radiation field).
Using equations (A2) and (28), we can write
$$\eqalign{
{{d T_\gamma}\over {dt}}
& = {{d T_\gamma}\over {d\tau}} {{d \tau}\over {d \mu}} {\dot \mu}_c =
- k_{es} \rho b \Gamma {\dot \mu}_c {{d T_\gamma}\over {d\tau}} \cr
& = - a c \Gamma v_c
{{2 k_{es} \rho Y_c}\over \tau} \Tg \left( {\Tg\over T} - 1 \right)
\, , \cr} \eqno({\rm A}3) $$
where $ {\dot \mu}_c = d\mu_c/dt $ and $v_c = b {\dot \mu}_c/ac
= (f + 1/3)^{1/2}$. However, also
$$ {{d T_\gamma}\over {dt}} = {{\partial \Tg}\over {\partial t}} \barra_\mu
+ {{\partial \Tg}\over {\partial \mu}} \barra_t
{\dot \mu}_c \eqno({\rm A}4) $$
and so we finally get
$$ {{\partial \Tg}\over {\partial t}} \barra_\mu = - a c \Gamma v_c
\left[ {{2 k_{es} \rho Y_c}\over \tau} \Tg \left( {\Tg\over T} - 1 \right)
+ {1\over {b\Gamma}}
{{\partial \Tg}\over {\partial \mu}} \barra_t \right] \, .
\eqno({\rm A}5) $$
In practice, we obviously cannot calculate the integrated value of
$\tau$ directly from expression (A1) evaluated along the
outward--pointing characteristics for the radiation since this
would involve knowledge of information ahead of the current time
reached in the calculation. Instead, we evaluated equation (A1) along
the time--slice and this should give reasonable values. While it is
important to calculate the {\it derivative} (A2) along the correct
directions in order to ensure a satisfactorily causal propagation of
information, the calculation of the {\it integral} (A1) should not be
so sensitive.

To derive the stationary limit of equation (A5), we need first to write it
in terms of the Eulerian time and radial coordinates $(\widetilde {t},r)$.
Using the chain rule for differentiation in equation (A5), we obtain
$$\eqalign{
{\widetilde {t}}_t {{\partial \Tg}\over {\partial {\widetilde {t}} }} \barra_r
+ r_t {{\partial \Tg}\over {\partial r}} \barra_{\widetilde {t}}
& = - a c \Gamma v_c \biggl[
{{2 k_{es} \rho Y_c}\over \tau} \Tg \left( {\Tg\over T} - 1 \right)
\biggr. \cr
& \left. + {1\over {b\Gamma}} {\widetilde {t}}_\mu
{{\partial \Tg}\over {\partial {\widetilde {t}} }} \barra_r
+ {1\over {b\Gamma}} r_\mu {{\partial \Tg}\over {\partial r}}
\barra_{\widetilde {t}} \, \right] \, . \cr} \eqno({\rm A}6) $$
The condition of stationarity is expressed with respect to the fixed
(constant $r$) Eulerian observer by
$${\partial\over {\partial {\widetilde {t}} }} \barra_r = 0 \, .
\eqno({\rm A}7)$$
Taking the stationary limit in equation (A6) and using $r_t = acu$ and
$r_\mu = b\Gamma$ finally yields
$$ \left( 1 + {u\over {\Gamma v_c}} \right)
{{\partial \Tg}\over {\partial r}} \barra_{\widetilde {t}} =
- {{2 k_{es} \rho Y_c}\over \tau} \Tg \left( {\Tg\over T} - 1 \right)
\, , \eqno({\rm A}8) $$
which corresponds to equation (31).

\bye